\documentclass[12pt]{article}
\usepackage{graphicx}
\newcommand{\bb}{\begin{equation}}
\newcommand{\ee}{\end{equation}}
\newcommand{\ba}{\begin{eqnarray}}
\newcommand{\ea}{\end{eqnarray}}

\begin{document}

\title{{\bf Classicality of Consciousness\\ in Quantum Darwinism}}

\author{
Don N. Page
\thanks{Internet address:
profdonpage@gmail.com}
\\
Department of Physics\\
4-183 CCIS\\
University of Alberta\\
Edmonton, Alberta T6G 2E1\\
Canada
}

\date{2022 December 4}

\maketitle
\large
\begin{abstract}
\baselineskip 20 pt

A simple toy model is proposed that would allow conscious perceptions to be either classical (perceptions of objects without large quantum uncertainties or variances) or highly quantum (e.g., having large variances in the perceived position within a single perception), and yet for which plausible quantum states exhibiting Quantum Darwinism would lead to much higher measures for the classical perceptions.

\end{abstract}

\normalsize

\baselineskip 21.1 pt

\newpage

\section{Introduction}

Since unitary quantum theory almost inevitably makes it the case that for most times, nearly all macroscopic objects have large uncertainties in their positions relative to other objects not rigidly attached to them, there has been the mystery of why our conscious perceptions are generally of such objects being perceived as having relatively precise relative locations.  In other words, although our universe certainly seems to be quantum, our conscious observations seem to be almost entirely classical.

Decoherence \cite{Zeh1970,Zeh1973,Z1981,Z1982,Z1984,JZ1985,Z1989,Z1991,Albrecht1992,ZHP1993,PHZ1993,Z1993,Anderson, PZ1993,Albrecht1993,ZP1994,Z1994,ZP1995,AZ1996,AZ1996b,APZ1997,Z1998a,HSZ1998,Z1998b,Z1998c,PZ1999,Z2003a,Z2003b,Schloss2004,DDZ2005,Schloss2007} and Quantum Darwinism \cite{Z2000,OPZ1,OPZ2,BZ1,BZ2,Z0,KTPA,BZ3,Z1,PR2009,ZQZ2009,LB,Durt,ZQZ2010,RZ2010,RZ2011,Z15,Z2,RZZ,Z3, Korbicz,Ball,BKGM,TYGDZ,MW,Z4,Ollivier,GTYDZ,SB,TZLS,Zw2022,GH,TADC,Z5} are properties of the quantum state and its evolution at least in a region of the universe sufficiently like ours, particularly in having a strong thermodynamic arrow of time and having subsystems that exist moderately stably for time periods long compared with the decoherence timescale that is often microscopically short.  They can explain why a system or object can be redundantly recorded, in a suitable basis called the {\it pointer basis} \cite{Z1981,Z1982}, by the environment, especially by the photons that scatter off the object.  Observers who intercept a small fraction of this environment gain access to the redundant information that the environment records and presumably can also record this information redundantly.

However, the quantum state and dynamics do not by themselves logically imply what (if any) conscious perceptions occur, or how much.  (I use conscious perception and sentient experience interchangeably, meaning all that one is consciously aware of at once.)  Presumably there are rules for getting the sentient experiences and their measure from the quantum state, but we do not yet know much about what these rules of `psycho-physical parallelism' might be.  These measures would be needed to use as substitutes, after being normalized, for the likelihood of a theory for the quantum state and for these rules in complete theories that are deterministic (e.g., similar to Everettian quantum theory that has no random collapse of the wavefunction) and do not really have any probabilities from true randomness \cite{P2006,P2017}.


Perhaps the simplest framework for the rules giving the measures of the sentient experiences would be that the unnormalized measures are the expectation values, in the quantum state of the universe, of a certain set of positive operators, one for each possible sentient experience, which I shall call the Awareness Operator corresponding to the sentient experience \cite{P1994a,P1994b,P1995a,P1995b,P1996,P2001,P2006,P2011,P2017,P2020}.
The Awareness Operators are not projection operators, do not all have the same norm, and do not sum to the unit operator as a POVM would, so the sum of their expectation values depends on the quantum state and is generically not unity.  Thus the measures directly given by their expectation values are, in general, unnormalized, though for any single particular quantum state, such as the quantum state of the universe, they can be normalized.  (Of course, the measures might be more complicated functionals of the quantum state, such as non-unit powers of expectation values or other nonlinear functionals, but expectations values, which are linear in the quantum state density matrix, appear to be the simplest possibility.  For the unnormalized measures to be truly linear in the quantum state density matrix, the Awareness Operators themselves should be independent of the quantum state.)

Let us make the plausible assumption that most of the contribution to the expectation values of the Awareness Operators come from conditions inside conscious beings.  For conscious humans, it seems virtually certain that most of the contribution comes from inside their brains.  Therefore, for brevity, I shall refer to ``brains'' as the locations whose conditions give the dominant contributions for the measure of conscious perceptions, without meaning to prejudice the question of what gives the dominant contributions over the entire universe.

For sighted humans, it seems that most (but certainly not all) of their conscious awareness is usually of visual images, which are mostly mediated by photons scattering off objects, entering the eyes, and exciting retinal neurons that then process and transmit the signals to the brain, where more processing occurs before the resulting state in some region of the brain presumably contributes directly to the expectation values of the Awareness Operators.  Quantum Darwinism can help explain why stable records of the effect of the photons give information about the angular and spectral distribution of the photons entering the eyes, and hence about the approximate location and colors of objects, rather than about superpositions of greatly different locations.

However, if the Awareness Operators are determined {\it a priori} by the laws of psycho-physical parallelism connecting consciousness with the quantum state  (which, as a physicist, I regard as laws of physics, even though they are not laws logically deducible from any of the laws of physics so far discovered, so to distinguish them from the presently known laws of physics, I call them the augmented laws of physics), they would not be determined by the conditions outside the brain.  (The Awareness Operators themselves would also not be determined by the conditions inside the brain, but the expectation values they produce would of course depend on the conditions inside the brain that are themselves highly influenced by the signals coming in from the outside, even though I am assuming that the conditions outside the brain do not contribute directly to the expectation values.)  So this raises the question of why our conscious perceptions seem to be of aspects of the external world that are determined by Quantum Darwinism.  It seems mysterious why the state-independent Awareness Operators should be well tuned to the information recorded redundantly by Quantum Darwinism.

Here I shall propose that indeed the Awareness Operators are not themselves tuned to the conditions redundantly recorded by Quantum Darwinism, but that instead there are many sets of Awareness Operators that are each tuned to different forms of multiple copies of information, and it is the set of Awareness Operators that are tuned to the form of multiple copies that actually exist in the brain that will receive the dominant expectation values and hence make the dominant contribution to consciousness.  There may indeed be some measure of conscious perceptions that do not correspond to classical components of the quantum state, but if that measure is sufficiently suppressed in comparison with the measure corresponding to classical components, that would be sufficient to explain why our observations are predominantly classical.

\section{Model for the Brain and Awareness Operators}

As discussed above, assume that most of the contributions to the expectation values of the Awareness Operators (which I am assuming give the unnormalized measures for the corresponding conscious perception, one Awareness Operator for each conscious perception) come from the conditions inside brains, and plausibly from conditions in certain subregions of brains, conditions that have low measures of occurring elsewhere.  For example, one might think of each Awareness Operator as being essentially an integral over spacetime of a localized projection operator onto a certain set of configurations that predominantly occur only inside certain regions of brains when they have the corresponding conscious perception.  (Each Awareness Operator thus does not have any fixed location but only picks up expectation value from regions that do have an element of one of the configurations corresponding to the localized projection operator that is integrated over the spacetime to give the full Awareness Operator.) 

Here let us model this region of the brain by a certain set of $N \gg 1$ qubits.  Although these qubits need not be spins or have any particular directional properties, I shall visualize them as if they were spin-half particles (with no positional degrees of freedom) that would each, if it were in a pure state, have its spin point in some direction given by a unit vector ${\bf n}$ on the Bloch sphere; such a state can be characterized as having the density operator $\rho_{\bf n} = |{\bf n}\rangle\langle{\bf n}|$, the rank-1 projection operator onto this pure state.  Then I can write the probability that such a pure state labeled by its spin direction ${\bf n}$ would be measured to have spin direction ${\bf m}$ at an angle $\theta$ from ${\bf n}$, with $\cos{\theta} = {\bf m}\cdot{\bf n}$, as 
\bb
P({\bf m},{\bf n}) = Tr(\rho_{\bf m}\rho_{\bf n}) = \cos^2{(\theta/2)} = (1/2)(1 + {\bf m}\cdot{\bf n}).
\label{prob}
\ee  

(Incidentally, if one visualized the spin-half particle in the pure state $\rho_{\bf n} = |{\bf n}\rangle\langle{\bf n}|$ as being a small opaque ball with the hemisphere surface within $90^\circ$ of the direction ${\bf n}$ painted white and the opposite hemisphere painted black, then this probability is the fraction of the total solid angle that is seen as being white from a long way away in the direction ${\bf m}$.)

I shall assume that these qubits occur mainly only inside certain regions of brains.  Let the projection operator ${\bf P}_k$ (before integrating over the spacetime) corresponding to the Awareness Operator ${\bf A}_k$ (with $1\leq k \leq M$ labeling which of the $M$ total Awareness Operators of this type is being considered) be the tensor product of $N$ projection operators for each of the spins to be ``up'' along the same direction, given by the unit vector ${\bf m}_k$, but with that direction varying with $k$ that labels the Awareness Operator ${\bf A}_k$.  If all of the qubits were pointing in the direction ${\bf n}$, then the expectation value of this projection operator would be $\langle{\bf P}_k\rangle = P({\bf m}_k,{\bf n})^N = [(1/2)(1 + {\bf m_k}\cdot{\bf n})]^N$.  If the state of the $N$ spins were a maximally mixed state, the expectation value of the total projection operator at the location of this set of $N$ qubits would be $\langle{\bf P}_k\rangle = 2^{-N}$, which I am assuming is very small, since I am taking $N \gg 1$.

Now suppose that Quantum Darwinism leads to the density matrix of the $N$ qubits to be essentially an incoherent sum of the tensor product of the $N$ spin projection operators that has all the spins pointing along a particular direction given by the unit vector ${\bf n}$, with nonnegative real coefficient $p$, and of the tensor product of the $N$ spin projection operators that has all the spins pointing in the opposite direction, given by the unit vector ${\bf -n}$, with nonnegative real coefficient $1-p$.  In other words, the mixed state of the $N$ qubits is an incoherent mixture with probability $p$ that all of the spins are pointed in the ${\bf n}$-direction and probability $1-p$ that all of the spins are pointing in the opposite direction, given by the following density operator,
\bb
\rho = p|{\bf n},{\bf n},\ldots\rangle\langle{\bf n},{\bf n},\ldots|+(1-p)|{-\bf n},{-\bf n},\ldots\rangle\langle{-\bf n},{-\bf n},\ldots|, 
\label{qubit-state}
\ee
with $N$ qubits in each ket and bra of this density operator.

\newpage

However, assume that the augmented laws of physics do not restrict the Awareness Operators ${\bf A}_k$ to have just one single direction for their projection operators, but rather that instead they correspond to a whole set of directions ${\bf m}_k$.  Then the expectation value of the Awareness Operator ${\bf A}_k$ would be proportional to the expectation value of its corresponding projection operator 
\bb
{\bf P}_k = |{\bf m}_k,{\bf m}_k,\ldots\rangle\langle{\bf m}_k,{\bf m}_k,\ldots|
\label{Pk}
\ee
for all of the qubits to be pointing in the direction ${\bf m_k}$, which is
\bb 
\langle{\bf P}_k\rangle \equiv Tr({\bf P}_k\rho) = p[(1/2)(1 + {\bf m}_k\cdot{\bf n})]^N + (1-p)[(1/2)(1 - {\bf m}_k\cdot{\bf n})]^N.
\label{Pexp}
\ee

For a complete set of laws of physics, one would not only need the quantum state of the universe and the Awareness Operators whose expectation values in that quantum state give the unnormalized measures of the corresponding conscious perceptions, but also the content of each perception.  How the content of the perception is related to the corresponding Awareness Operator is far beyond the scope of my ability to predict, but surely there is some strong correspondence.  Here I shall just assume that if the direction ${\bf m}_k$ corresponding to the projection operator ${\bf P}_k$ that is integrated over spacetime to give the Awareness Operator ${\bf A}_k$ is closely aligned with the direction ${\bf n}$ the spins would have if they corresponded to a classical pointer basis, then the content of the corresponding conscious perception would be an awareness of a classical perception, without large quantum uncertainties.  In other words, the conscious perception would include an awareness of an image that appears to be of one or more objects at definite directions from the observer.

Let us assume that an Awareness Operator ${\bf A}_k$ with $N$-spin product projection operator ${\bf P}_k$ having a direction ${\bf m}_k$ close to that of ${\bf n}$ would correspond to a classical awareness of the visual image recorded redundantly by the $N$ qubits all pointing in the direction ${\bf n}$ (with probability $p$, there also being probability $1-p$ that the $N$ qubits are all pointing in the opposite direction, ${-\bf n}$, which I am assuming would lead to a complementary classical awareness, so that an ${\bf A}_k$ with ${\bf m}_k$ close to that of ${-\bf n}$ would contribute significant measure for the complementary classical awareness).  By ${\bf m}_k$ being ``close'' to ${\bf n}$ or to ${-\bf n}$, I mean that either ${\bf n}\cdot{\bf m}_k$ or ${-\bf n}\cdot{\bf m}_k$ is within $f$ of unity for $f\ll 1$, i.e., that $1-f \leq |{\bf n}\cdot{\bf m}_k| \leq 1$.  If there are many ${\bf m}_k$'s uniformly distributed over the unit sphere, $f$ is the fraction of them that are ``close'' to ${\bf n}$ or to ${-\bf n}$ in this sense.  This implies that if the brain qubits corresponded to directions distributed uniformly over the unit sphere, only a fraction $f$ of the total measure for all the Awareness Operators of this form would correspond to classical awarenesses.

\section{Fraction of the Measure That Is Classical}

If the $N$ relevant brain qubits are in the mixed state given by Eq.\ (\ref{qubit-state}),
as plausibly given by Quantum Darwinism, then the situation is quite different from having brain qubits distributed randomly in direction.  If $c = \cos{\theta} = {\bf n}\cdot{\bf m}_k$ with $\theta$ being the angle between ${\bf n}$ and ${\bf m}_k$, then the ratio of the contribution of the fraction $f$ of the Awareness Operator directions ${\bf m}_k$ (assuming that they are uniformly distributed over the unit sphere) that are ``close'' to ${\bf n}$ or ${-\bf n}$ (have $1-f\leq |c|\leq 1$ and hence are assumed to give the measure for a classical awareness), to the total measure given by all the Awareness Operators is, with $\Theta$ being the Heaviside step function,
\ba
F&\equiv&\frac{\int_{-1}^1 dc\, \Theta(|c|-1+f)\langle{\bf P}_k\rangle}{\int_{-1}^1 dc \langle{\bf P}_k\rangle}
=\frac{\int_{-1}^1 dc\, \Theta(|c|-1+f)\left[p\left(\frac{1+c}{2}\right)^N+(1-p)\left(\frac{1-c}{2}\right)^N\right]}
{\int_{-1}^1 dc\, \left[p\left(\frac{1+c}{2}\right)^N+(1-p)\left(\frac{1-c}{2}\right)^N\right]}\nonumber \\
&=& 1-\left(1-\frac{1}{2}f\right)^{N+1}+\left(\frac{1}{2}f\right)^{N+1}\equiv 1-\delta.
\label{fraction}
\ea
Therefore, all but a fraction $\delta = 1-F = (1-f/2)^{N+1}-(f/2)^{N+1}$ of the measure for this type of conscious perceptions will be perceived as classical in this toy model.  For a very small fraction $f$ of directions for the Awareness Operator qubit directions ${\bf m}_k$ and for large $N$, 
\bb
\delta \approx (1-f/2)^{N+1} = e^{(N+1)\ln{(1-f/2)}} \approx e^{-\frac{N+1}{2}f}.
\label{delta}
\ee
This implies that if $f\ll 1$, to get at least all but a fraction $\delta$ of the measure of classical conscious perceptions to be classical, the number of qubits $N$ in the projection operator ${\bf P}_k$ that is integrated over the spacetime to give the Awareness Operator ${\bf A}_k$ must be at least $N\approx (2/f)\ln{(1/\delta)}-1$.  Alternatively, the fraction of the solid angle $f$ given by the directions ${\bf m}_k$ that make classical perceptions have a measure that is at least a fraction $F = 1-\delta$ of the total must be at least $f \approx [2/(N+1)]\ln{(1/\delta)}$.  Since $\ln{(1/\delta)}$ is only logarithmically large, it is not too hard to get $N$ large enough that the solid angle fraction $f$ can be small, even for small $\delta$ that is the fraction of the measure of these conscious perceptions that are not classical.

\section{Conclusions}

Awareness Operators, which are to be specified by augmented laws of physics independently of the quantum state, are postulated to give the unnormalized measures of the corresponding conscious perceptions.  However, they can have a form in which most of the measure is for conscious perceptions that are classical, if consciousness occurs primarily in a region where decoherence and Quantum Darwinism lead to high redundancy of information.
One does need to assume that the fraction of the Awareness Operators that get large expectation values from quantum states of the brain in which information redundantly by Quantum Darwinism is not too small, though this fraction can be considerably smaller than unity if the information is stored sufficiently redundantly.  One also needs to assume that there is a strong correspondence between the content of conscious perceptions and the quantum states that produce the measures of these perceptions as the expectation values of the corresponding Awareness Operators.



\section{Acknowledgments}

I am grateful especially for email discussions with James Hartle and Wojciech Zurek, which motivated this analysis.  My research was funded by the Natural Sciences and Engineering Research Council of Canada.

\end{document}